\documentclass[12pt,preprint]{aastex}
\usepackage{mathptmx}
\usepackage{natbib}
\usepackage{epsf}
\usepackage{colordvi}

\shorttitle{The Formation and Eruption of A Sigmoidal Filament Driven by Rotating Network Magnetic Fields }
\shortauthors{Dai et al.}

\begin{document}

\title{The Formation and Eruption of A Sigmoidal Filament Driven by Rotating Network Magnetic Fields }

\author{Jun Dai\altaffilmark{1,2}, Haisheng Ji\altaffilmark{1,2}, Leping Li\altaffilmark{3,4}, Jun Zhang\altaffilmark{3,4}, Huadong Chen\altaffilmark{3,4}}

\affil{$^1$Key Laboratory of Dark Matter and Space Astronomy, Purple Mountain Observatory, CAS, Nanjing, 210008,China}
\affil{$^2$School of Astronomy and Space Science, University of Science and Technology of China, Hefei, 230026,China}
\affil{$^3$Key Laboratory of Solar Activity, National Astronomical Observatories, Chinese Academy of Sciences, Beijing 100101, China}
\affil{$^4$University of Chinese Academy of Sciences, Beijing, 100049, China}
\email{}

\begin{abstract}
We present the formation and eruption of a sigmoidal filament driven by rotating network magnetic fields (RNFs) near the center of the solar disk, which was observed by the one-meter aperture New Vacuum Solar Telescope (NVST) at Fuxian Solar Observatory (FSO) on 2018 July 12. Counterclockwise RNFs twist two small-scale filaments at their northeastern foot-point region, giving a rotation of nearly $200^\circ$ within about 140 minutes. The motion of the RNF has a tendency to accelerate at first and then decelerate obviously, as the average rotation speed increased from $10^\circ$ hr$^{-1}$ to $150^\circ$ hr$^{-1}$, and then slowed down to $50^\circ$ hr$^{-1}$. Coalescence then occurs between filments F1 and F2. Meanwhile the fine structures in the southwestern region of the filament was involved in another interaction of coalescence. The subsequent EUV brightening due to plasma heating is observed in the two interaction regions. These interacting structures, including F1, F2 and the fine structures in the southwestern region, eventually evolve into a larger-scale sigmoidal filament twisted in the same direction as the RNFs gave. The twist of the sigmoidal filament has exceeded $4\pi$ and the filament erupted finally. The motion of the sigmoidal filament keeps uniform until a nearby jet collides, causing the filament to erupt faster. These results provide evidence that RNF plays an important role in the formation and eruption of the sigmoidal filament. The phenomena also suggests that the kink instability is the trigger mechanism for the filament eruption.

\end{abstract}

\keywords{Sun: activity - Sun: filaments, prominences - Sun: magnetic fields - Sun:Chromosphere}

\section{Introduction}

In recent years, some new observations have found that small-scale vortex flows in the photosphere could drive the corresponding rotating magnetic structure in the corona. Classically, the extreme ultraviolet (EUV) cyclones discovered by \cite{2011ApJ...741L...7Z} and magnetic tornadoes reported by \cite{2012Natur.486..505W}
are all driven by the underlying photospheric vortex flows or rotating network magnetic fields (RNFs), since the EUV and magnetic tornadoes are found to be rooted in the network magnetic fields.

In fact, similar theories have been proposed earlier: the motion of the photosphere may cause the pre-existing magnetic field of the corona to recombine and form a filament \citep[]{1989ApJ...343..971V,1997ApJ...479..448G}. Recent high resolution observations also hold that the rotating magnetic fields are closely associated with the formation and eruption of sigmoidal filaments \citep[]{2012ApJ...754...16Y,2013SoPh..286..453T,2014AAS...22430303C}. By studying the positions of the barbs of the filaments, \cite{2005SoPh..227..283L} concluded that the quiescent filaments are basically connected with the network magnetic fields below. Furthermore, \cite{2015ApJ...803...86Y} directly observed the formation and eruption of a quiescent circular filament driven by a RNF, which suggested that the RNFs will influence the evolution of the filaments in quiet regions.

Sigmoids are the manifestation of highly sheared magnetic structures, which present the S-shaped or inverse-Shaped structures in the lower corona \citep[]{1996ApJ...464L.199R}. According to a statistical research work \citep[]{1998GeoRL..25.2481H,1999GeoRL..26..627C,2007ApJ...671L..81C}, such sigmoidal structures are basically unstable and will erupt eventually. A qualitative model for sigmoid eruption was proposed by \cite{1980IAUS...91..207M} and further elaborated by \cite{2001ApJ...552..833M}, they suggested that a magnetic explosion is unleashed internally in the middle of the sigmoid. The prediction of Moore's model in observation includes the brightenings in multi-wavebands \citep[]{2006SoPh..235..147Y}, hot moving plasma \citep[]{2014ApJ...795....4J} and bright plasma outflows \citep[]{2014ApJ...797L..15C,2016ApJ...818L..27C} in the merging region of two filaments.

\cite{2000ApJ...540L.115C} reported that filaments sometimes appears as mixtures of threads based on high-resolution observations, and presented that we can discern the crossing of the overlying and underlying threads to determine the twist of a filament. It was found that the sigmoids almost have a twist less than $2\pi$ \cite[]{2003ApJ...596L.255L}. The twist ($\Phi$) is the rotating angle of the magnetic field lines through a flux rope, and continuous twisting around the flux rope inevitably leads to kink instability, which is a possible mechanism for solar filament eruption. \cite{2003A&A...406.1043T} found that the critical threshold for the twist lies in the range $2.5\pi<\Phi_{c}<2.75\pi$ by simulating a flux tube driven by photospheric vortex motions. Considering the effect of overlying magnetic field in the process of simulation, \cite{2005ApJ...630L..97T} further found that the decrease of the overlying magnetic field with height will decide whether the instability leads to a CME. If the magnetic field above the flux rope is strong enough, it would prevent the eruption of the flux rope, resulting in a failed eruption of filaments as reported by \cite{2003ApJ...595L.135J}.

On 2018 July 12, the formation and eruption of a small-scale sigmoidal filament in a quiet region near the center of the solar disk was recorded by NVST. In this paper, we focus on the rotation of the RNFs in the northeastern filament foot-point region and its role in the formation and twsiting of a sigmoidal filament. They provide us a rare opportunity to study the relationship between RNFs and sigmoidal filaments. The observations and results are separately presented in Sections 2 and 3. Our summary and discussion are presented in Section 4.

\section{Data and observations}

NVST \citep[]{2014RAA....14..705L} is a one-meter aperture ground-based vacuum solar telescope at Fuxian Solar Observatory, which is operated by Yunnan Observatories of the Chinese Academy of Sciences. Currently, it provides Level-1 high-resolution images of chromosphere and photosphere at H$\alpha$ line (sometimes including off-band) and TiO 7058\,\AA\ . Quick-look movies can be downloaded from its website http://fso.ynao.ac.cn/datacrvhive-ql.aspx. On 2018 July 12, There was no active region on solar disk. NVST pointed to the center of the solar disk with targetting a large quiescent filament. The active filaments analyzed in this paper is located north-east from the large quiescent filament. In the study, we employ NVST data obtained in the H$\alpha$ 6562.8\,\AA\ line from 05:23 UT to 10:16 UT. The H$\alpha$ images have a spatial sampling and time cadence of 0.137\arcsec\,pixel$^{-1}$ and 12\,s, respectively.

 For other wavelengths, we use EUV observations at 171\,\AA\ and 304\,\AA\ from Atmospheric Imaging Assembly \citep[AIA,][]{lemen12} on board Solar Dynamic Observatory \citep[SDO,][]{pesn12}. The emission of AIA 171 \,\AA\ and 304 \,\AA\ are from of Fe IX and He II, with the characteristic temperatures of $6.3 \times 10^5$ K and $5.0 \times 10^4$ K respectively. The large quiescent filament was recognizable at the two wavelength, which is very helpful for a believable co-alignment between EUV and H$\alpha$ images. Since the pixel resolutions of the two kind of images have been fixed, the only thing left for the co-alignment is to make the filament to be co-spatial. For level 1.5 AIA images, their spatial sampling is fixed at 0.6\arcsec\,pixel$^{-1}$. Visual inspections were made during the co-alignment, and we estimate the accuracy to be within 1\arcsec\,.

To analyze the nature of the RNF, we use network magnetic elements at the foot-point region of the active filaments observed by line-of-sight magnetograms of Helioseimic and Magnetic Imager \citep[HMI,][]{schou12} also onboard SDO. The 0.5\arcsec\,pixel$^{-1}$ and 45 s cadence of the magnetograms are sufficient to resolve the two network magnetic elements with the size of $\sim$ 5 arcsec and the temporal behavior of the rotation.

\section{Results}

Near the center of the solar disk, an unstable sigmoid filament was formed in H$\alpha$ 6562.8\,\AA\  channel and AIA 171\,\AA\ channel around 09:05 UT on 2018 July 12. It is denoted by white dotted lines in Figures 1((a)-(b)). We notice that the filament is located in a region with network magnetic fields in quiet Sun. The corresponding photospheric magnetic fields are given in Figure 1(c). The filament connects the southwestern positive and northeastern negative polarity network magnetic fields separately. No magnetic polarity inversion line (PIL) can be clearly discerned along the filament.

 At about 05:30 UT, the network magnetic field in the northeastern foot-point region of the filament starts to rotate slowly. In Figure 2(a), a white line is used to connect the centers of two concentrations of network magnetic field, and a red vertical line is drawn as a reference line. The angle $\alpha$ between the white line and the red line is used to record the rotation process of the RNFs. In Figure 2, we use the green and blue arrows to mark the negative network magnetic fields N1 and N2, respectively, and the arrows are made to be parallel to the line connecting N1 and N2, i.e., the white line in Figure 2(a). It can be seen from Figure 2 that the relative position and size of the negative RNFs do not change significantly, and the relatively independent topological structure of the negative RNFs is almost maintained from 05:30 UT to 08:00 UT. After this time period, as shown in Figure 2(h), N1 and N2 finally merged becoming one single negative network magnetic field.

Figure 3 shows the temporal evolution of the rotation angle $\alpha$ of the RNFs in the northeastern foot-point region of the filament. From 05:30 UT to 07:50 UT, the RNFs rotates counterclockwise by nearly $200^\circ$. From Figure 3, we find that the rotation has an obvious process of an acceleration, keeping nearby uniform and a deceleration. From 05:30 UT to 06:15 UT, the average angular velocity of rotation, V1, is  $10^\circ$ hr$^{-1}$. In a process of acceleration, from 06:30 UT to 07:15 UT, the average angular velocity of rotation, V2, becomes $150^\circ$ hr$^{-1}$, and then the rotation starts to slow down, from 07:25 UT to 07:50 UT, with an the average rotation velocity, V3, of $50^\circ$ hr$^{-1}$. At about 08:00 UT, RNFs stop their rotation, then the negative network magnetic fields N1 and N2 begin to merge, (see the animation of Figure 2), so the angle $\alpha$ cannot be measured any more. We use second-order piecewise fitting to get the blue curve value in Figure 3, which is a good approach to fit the variation of the rotation angles with time.

Starting around 07:30 UT, two small filaments F1 and F2 are observed in the H$\alpha$ 6562.8\,\AA\ band. We find that, at the moment when the filaments F1 and F2 are first observed, no obvious S-shaped structures can be detected. In the case of the rotary drive of RNFs, from around 07:50 UT, filaments F1 and F2 begin to bend into S-shaped structures (Figure 4(a-c)). In Figures 4(a)-(c), we show three snapshots of the structures of F1 and F2, and outline them with white dotted lines, the black contours represent the RNFs at the negative polarity foot-point of the filament F1 and F2. These prove that the F1 and F2 are rooted in the RNFs. After about 10 minutes, by around 08:00 UT, the filaments F1 and F2 are fully developed into S-shaped structures. From about 08:00 UT, almost the same time that RNFs stop the rotation, coalescence between the two filaments F1 and F2 takes place, the position of the interaction was marked by the blue arrows in Figures 4(d)-(e). Significant brightening is observed at the same location in AIA 171\,\AA\ (designated as B1 in Figure 4(f)). At the same time, in the region R2 enclosed by the blue dotted line in Figures 4(c)-(e), coalescence also occurs between many small-scale fine structures in the southwestern region, and significant brightening is also observed in the same region in AIA 171\,\AA\, enclosed by the pink dotted line in Figure 4(f).

Around 08:25 UT, a series of multiple coalescence occurred and resulted in a larger sigmoidal filament. In the process of the formation, the filament becomes unstable and shows a dynamic behavior. According to the identification method with observation \citep[]{2000ApJ...540L.115C}, we estimate the twist of the filament. As shown in Figure 4(g), two dark threads cross each other four times, which are pointed by four white arrows, indicating that the twist of the filament has exceeded $4\pi$, Which has exceeded the critical value of the kink instability triggering mechanism \citep[]{2003A&A...406.1043T}. At this point, the filament becomes unstable, begin to show a dynamic behavior. The twisting direction is same as the direction of RNFs. After about 15 minutes of the slow evolution, at around 08:40 UT, the northeastern foot-point of the filament shows a vortex-like structure, and the spine of the filament F shows an inverse sigmoidal shape. (see the animation of Figure 4). The vortex-like structure and inverse s-shaped structure as shown in Figure 4(h) are well developed at 08:56 UT.

The motion of the filament keeps uniform until about 09:03 UT, a jet collides with the moving filament F. At the interface, brightening is identified, see Figures 5(b)-(c), which lasts for about three minutes until 09:12 UT. We make a time slice of the H$\alpha$ 6562.8\,\AA\ images, and the space time diagram is shown in Figure 5(e). It indicates that the filament first moves slowly with a rough uniform speed across the field-of-view of ~3 km s$^{-1}$, and then quickly appears to erupt with a larger mean projected speed of ~12 km s$^{-1}$. No associated coronal mass ejection is detected during the filament eruption. The filament eruption is thus a failed one.

\section{Summary and discussion}

In this paper, we investigated the formation and eruption of a sigmoidal filament observed by NVST on 2018 July 12. The formation and eruption of the filament are associated with the magnetic rotating network fields (RNFs). By analyzing the evolutions of the RNFs in the northeastern foot-point region of the filaments, we notice that the RNFs results in the bending of two small-scale filaments, F1 and F2. Due to the persistent twisting of RNFs, the coalescence between the two filaments and also the fine structures in the southwestern region occurred, forming a larger sigmoidal filament. At the same time, the twist of the sigmoidal filament increased to at least $4\pi$ and became unstable. What characterizes the instability is the subsequent slow movement of the filament, which keeps nearly uniform until accelerated by a nearby jet. According to the kink instability condition, we suggest that filament eruption studied in the paper was be triggered by kink instability.

 S-shaped filaments caused by RNFs have been reported by \cite{2015ApJ...803...86Y}, who found that RNFs drove forming a small circular filament in the quiet Sun. Similarly, sunspot rotation is closely associated with the formation of the sigmoidal filaments \citep[]{2012ApJ...754...16Y,2013SoPh..286..453T,2014AAS...22430303C}. Our results provide further observational evidence that rotating motion on various spatial scales in the photosphere plays an important role in the formation of the quiescent filaments.

 In this event, the RNFs rotated by $200^\circ$ within about 140 minutes, which was less than the twist value $4\pi$. We speculate that the multiple coalescence between two filaments F1 and F2 and the fine structures in the southwestern region directly lead to an increase in the twist of the filament. Corresponding brightening in the EUV band is identified at the regions where the coalescence takes place, in agreement with the observations previously reported, in which the brightening plasma structures in multi-wavebands \citep[]{2006SoPh..235..147Y,2014ApJ...795....4J,2014ApJ...797L..15C,2016ApJ...818L..27C} in the merging region of two filaments are observed. In our study, RNFs and the coalescence twist the quiescent filament together. The joint twisting is rarely reported.

\cite{2003A&A...406.1043T} confirmed that kink instability is one of the triggering mechanism for solar filament eruptions, which was later supported by some observational studies \citep[]{2010ApJ...715..292S,2012ApJ...746...67K,2014ApJ...782...67Y,2019BlgAJ..30...99D}. However, the twist of the filament is not easy measure. In this event, based on the high resolution images, we clearly distinguish the crossing threads structure in the spine of the filament. The twist of the filament observed in this paper has reached $4\pi$, exceeding the critical value of the kink instability obtained by previous authors \citep[]{2003A&A...406.1043T}.

The observation and concept of failed filament eruption was first proposed in \cite{2003ApJ...595L.135J}, \cite{2005ApJ...630L..97T} obtained the same conclusion through numerical simulation: strong overlying magnetic field lines will inhibit the lifting of the filament and lead to a failed eruption. Later similar failed eruptions was reported \citep[]{2006ApJ...653..719A,2009ApJ...696L..70L,2013A&A...552A..55K,2016PASJ...68....7X}.The filament eruption in this work should be failed since no associated coronal mass ejection is detected. In addition, \cite{2011RAA....11..594S} considered that the amount of energy gained during the eruption is also an important determinant of whether the eruption is full or not. The filament eruption reported in this work is actually a gentle one. We think that because the structural scale of the filament is too small, therefore, a strong enough overlying magnetic field and insufficient energy obtained during the eruption together lead to the failed eruption.

The authors thank the referee for constructive comments and suggestions. The H$\alpha$ data used in this paper were obtained with the New Vacuum Solar Telescope in Fuxian Solar Observatory of Yunnan Astronomical Observatory, CAS. SDO is a mission of NASA's Living With a Star Program. AIA and HMI data are courtesy of the NASA/SDO science teams. The authors are indebted to the SDO and NVST teams for providing the data. H.J. is supported by the National Natural Science Foundations of China (11790302), L.L. is supported by the National Natural Science Foundations of China (11673034 and 11533008). J.Z. is supported by the National Natural Science Foundations of China (11790304), and Key Programs of the Chinese Academy of
Sciences (QYZDJ-SSW-SLH050)

\begin{figure}
\includegraphics[width=1. \textwidth]{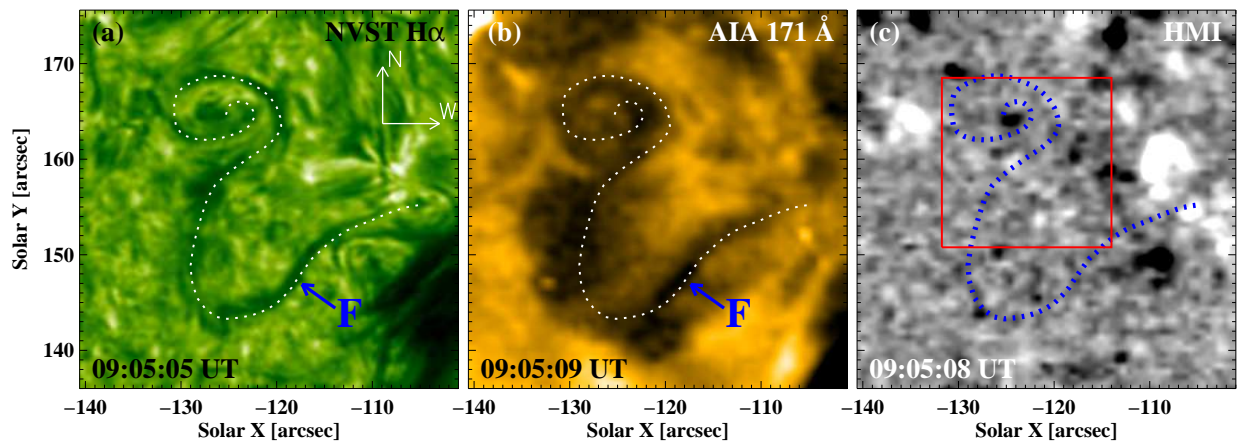}
\caption{\small{General information of the filament. (a) A NVST H$\alpha$ image. (b) An AIA 171 \AA~image. (c) An HMI LOS magnetogram with the magnetic field strengths in the range of -15 and 15 G.  The white and blue dotted lines in ((a)-(c)) outline the filament F in all the panels. The red rectangle in (c) shows the field of view (FOV) of Figure\,\ref{fig2}.}}
\label{fig1}
\end{figure}

\begin{figure}
\includegraphics[width=1.\textwidth]{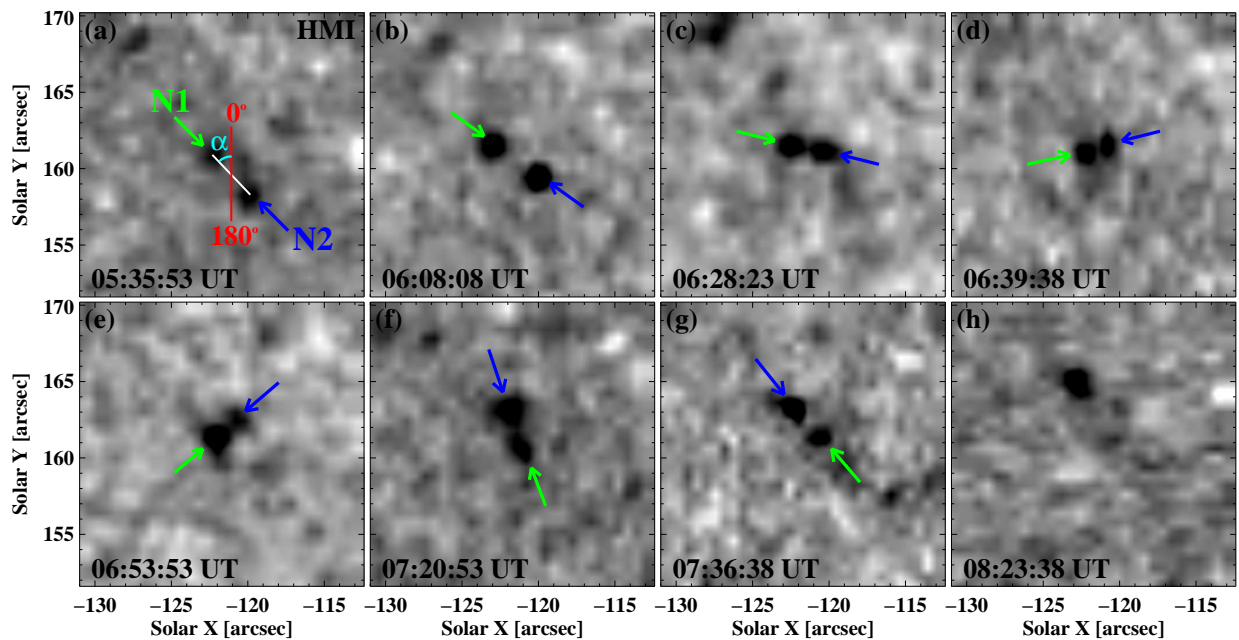}
\caption{\small{Evolution of magnetic fields at the northeastern endpoint of the filament. (a-h) HMI LOS magnetograms. The white line in (a) indicates the connection between the centers of the two negative network magnetic fields, N1 and N2, and the red line in (a) marks 0 degrees or 180 degrees. $\alpha$ in (a) is the angle between the white and red lines.  The green and blue arrows in (a-g) points to the negative network magnetic fields N1 and N2.(An animation of this figure is available.)}}
\label{fig2}
\end{figure}

\begin{figure}
\includegraphics[width=1.\textwidth]{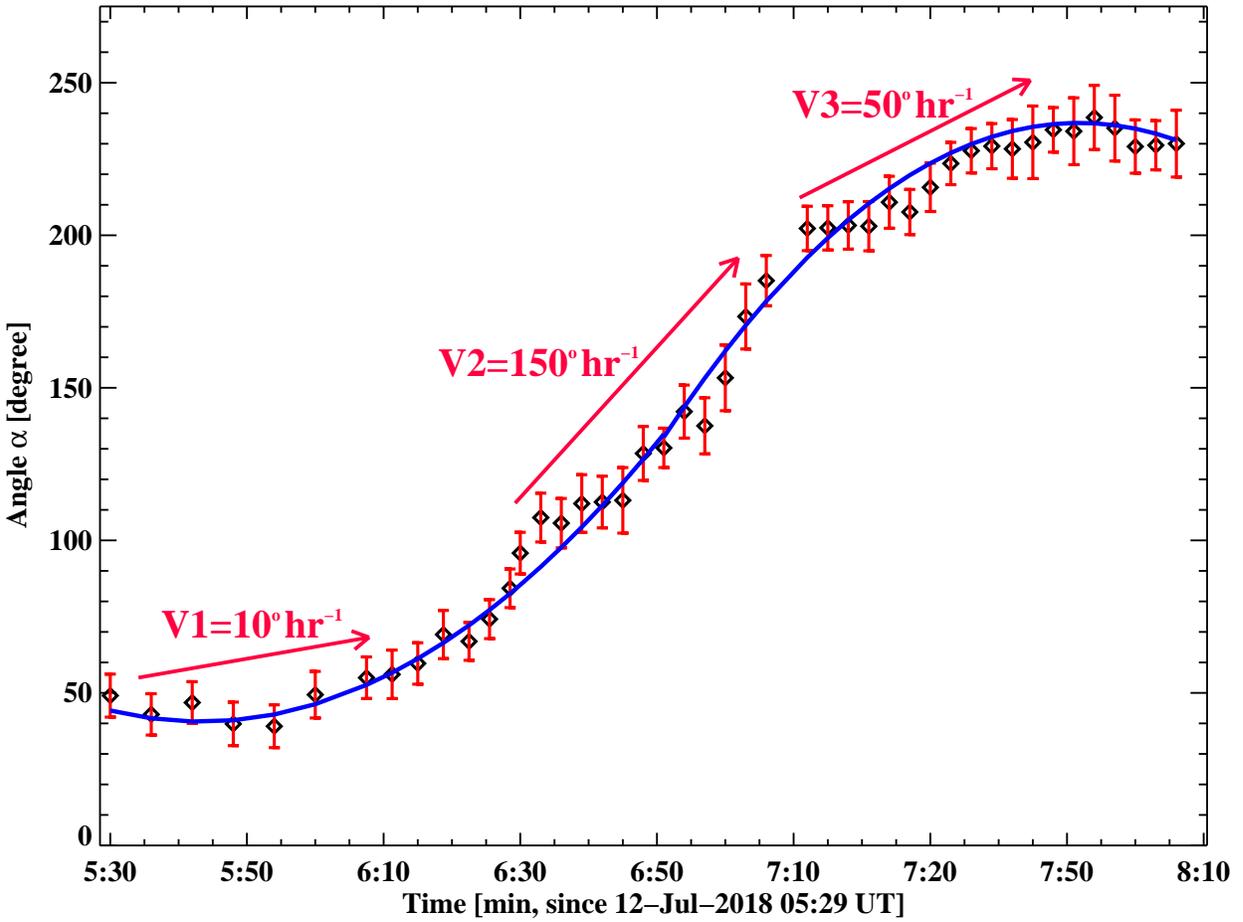}
\caption{\small{Temporal behavior for the rotation of the magnetic fields at the northeastern endpoint of the filament. The black squares correspond to the rotating Angle $\alpha$ (the average angle of several measurements) in Figures\,\ref{fig2}. The red bars mark the errors (the difference between the maximum or minimum value measured and the average value). The blue line represents a fit to the observed values and shows its evolution trend. The red arrows represent the rotations in three different time periods, and the rotating speeds are marked by the numbers.}}
\label{fig3}
\end{figure}

\begin{figure}
\includegraphics[width=1.\textwidth]{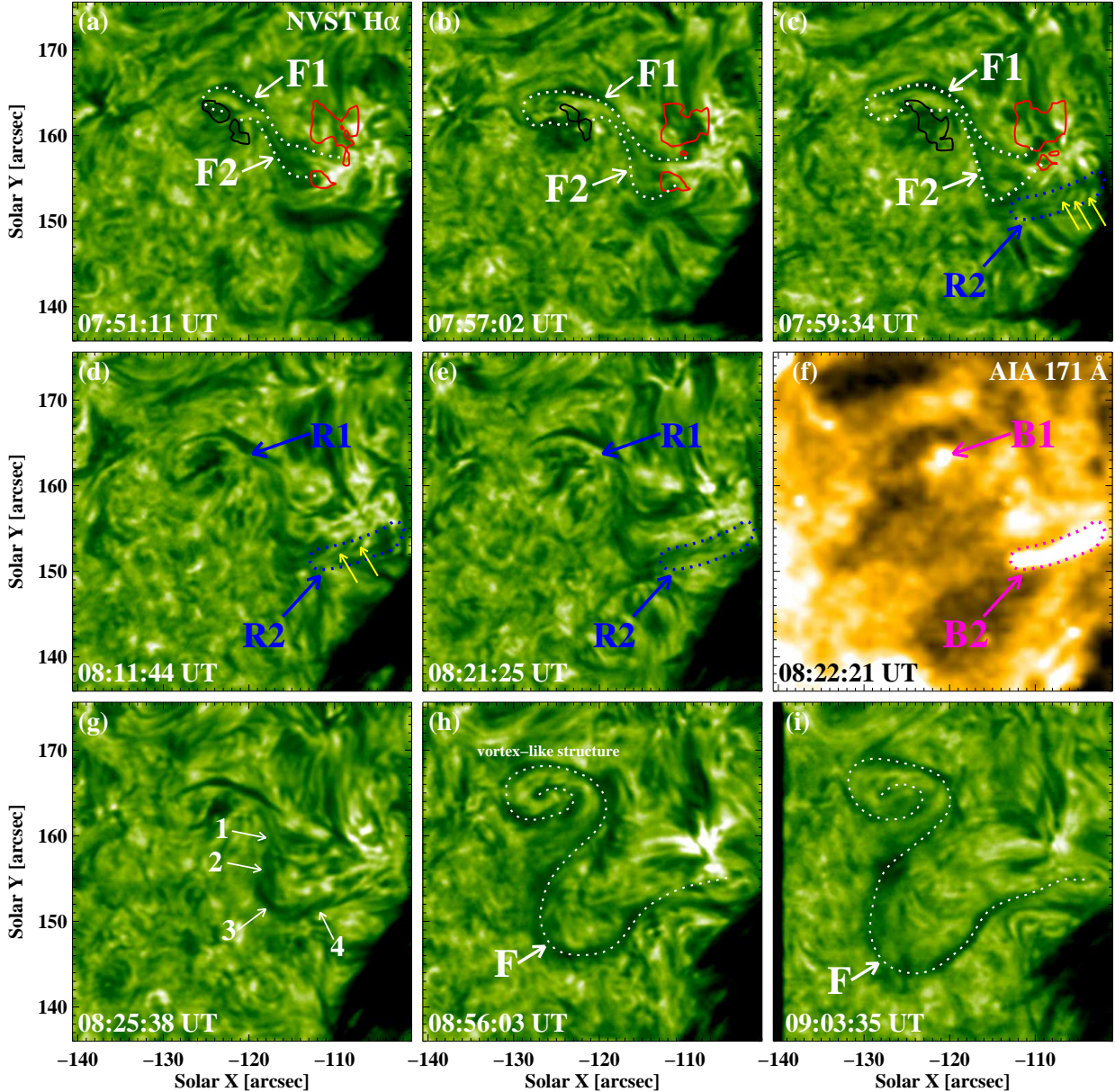}
\caption{\small{Formation of the sigmoidal filament. All panels except panel (f) give NVST H$\alpha$ images. Panels (f) gives an AIA 171 \AA~image to show the brightenings during the coalescense. The white dotted lines in ((a)-(c)) outline the initial sigmoid structure of the filament F1 and F2 at different times, respectively. The black and red contours in ((a)-(c)) represent the negative and positive polarity magnetic fields surrounding the filament F1 and F2 with the magnetic field strengths of -15 and 15 G, respectively. The small yellow arrows in ((c)-(d)) mark the fine structures also being involved in the coalescence. The blue arrows in ((d)-(e)) mark the coalescence interaction region R1 between F1 and F2. The blue dotted circles in ((c)-(e)) indicate the coalescence interaction region R2 between F2 and fine structures. The pink arrow in (f) marks a brightening region B1. The pink dotted circle in (f) marks a brightening region B2. The white arrows, 1-4, in (g) mark the crossing positions of two dark threads in the filament F. The white dotted lines in ((h)-(i)) outline the filament F. (An animation of this figure is available.)}}
\label{fig4}
\end{figure}

\begin{figure}
\includegraphics[width=0.8\textwidth]{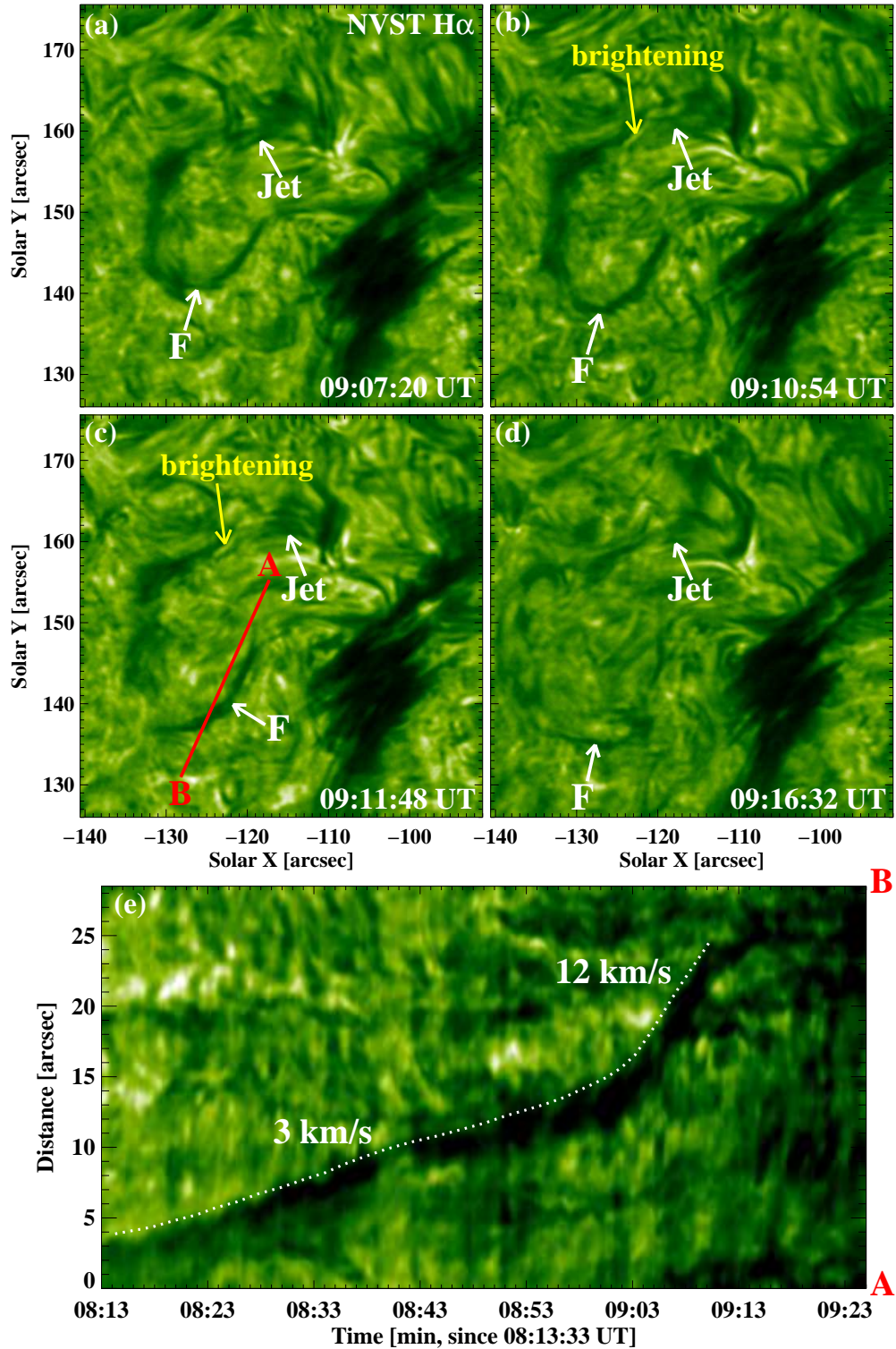}
\caption{\small{Panels (a-d) give the eruption process of the filament as observed with  NVST H$\alpha$ images. (e) A time-slice of NVST H$\alpha$ images along the red line AB as displayed in (c). The white dotted line in (e) outlines the eruption of the filament F, and the erupting speeds are denoted by the numbers in (e). (An animation of this figure is available.)}}
\label{fig5}
\end{figure}

\end{document}